\documentclass[english]{webofc}
\usepackage[varg]{txfonts}

\newcommand{\thcm}{\theta_{\textrm{c.m.}}}
\newcommand{\sigexp}{\sigma_{\textrm{exp}}}
\newcommand{\bra}[1]{\langle #1 |}
\newcommand{\ket}[1]{| #1 \rangle}
\renewcommand{\Sigma}{\varSigma}

\woctitle{MENU 2013 -- 13th International Conference on Meson-Nucleon Physics and the Structure of the Nucleon}

\begin{document}
\title{Complete sets of observables in pseudoscalar-meson photo\-production}

\author{Tom Vrancx\inst{1}\fnsep\thanks{\email{tom.vrancx@ugent.be}} \and
        Jan Ryckebusch\inst{1} \and
        Jannes Nys\inst{1}
}

\institute{Department of Physics and Astronomy, Ghent University, Proeftuinstraat 86, B-9000 Gent, Belgium}

\abstract{Pseudoscalar-meson photoproduction is characterized by four complex reaction amplitudes. A complete set is a minimum theoretical set of observables that allow to determine these amplitudes unambiguously. It is studied whether complete sets remain complete when experimental uncertainty is involved. To this end, data from the GRAAL Collaboration and simulated data from a realistic model, both for the $\gamma p \to K^+ \Lambda$ reaction, are analyzed in the transversity representation of the reaction amplitudes. It is found that only the moduli of the transversity amplitudes can be determined without ambiguity but not the relative phases.}
\maketitle
\section{Introduction}
\label{sec:introduction}
Quantum mechanics dictates that observables can be expressed as bilinear combinations of a certain number of complex amplitudes. An example of pseudoscalar-meson photoproduction is the $\gamma p \to K^+ \Lambda$ reaction. In meson photoproduction, two kinematic variables are involved: for example the invariant mass $W$ and the meson scattering angle in the center-of-mass frame $\thcm$. Since the photon, the target and the recoil particle each have two possible spin states, four independent complex amplitudes can be distinguished, considering the conservation of angular momentum. Since quantum states are only determined up to a constant phase factor, all information about the reaction is contained in four moduli and three relative phases.

A set containing a minimum number of observables from which the moduli and relative phases of the complex amplitudes can be determined (unambiguously) is dubbed a \emph{complete set}. In Ref.~\cite{Chiang:1996em} it was pointed out that theoretical complete sets consist of eight well-chosen observables, contrary to the nine observables suggested in Ref.~\cite{Barker:1975bp}. Obviously, observables cannot be determined with infinite precision. It is now the question whether or not eight observables still suffice to reach a complete knowledge about the four complex amplitudes?

\section{Complete sets in the transversity basis}
\label{sec:transversity_basis}
The transversity amplitudes in pseudoscalar-meson photoproduction are defined as $b_1 = {}_y\bra{+} J_y \ket{+}_y$, $b_2 = {}_y\bra{-} J_y \ket{-}_y$, $b_3 = {}_y\bra{+} J_x \ket{-}_y$, and $b_4 = {}_y\bra{-} J_x \ket{+}_y$. Here, ${}_y\bra{\pm}$ and $\ket{\pm}_y$ represent the spinors for the recoil and the target particle, and $J_{x,y}$ is the photon current. The subscripts indicate the polarization direction of the particle in question. The normalized transversity amplitudes are defined as $a_j = b_j/\sqrt{\sum_{i = 1}^4|b_i|^2}\equiv r_j e^{i\alpha_j}$, with $r_i$ and $\alpha_i$ the moduli and the phase of the amplitude $a_i$.In order to determine the normalized transversity amplitudes $a_i$,  information about the unpolarized differential cross section is no longer needed and complete sets are reduced from eight to seven observables.

In Table I of Ref.~\cite{Vrancx:2013pza}, the expressions for the three single and twelve double asymmetries in the transversity representation are listed. The moduli $r_i$ can be readily expressed in terms of the single asymmetries $\Sigma$, $T$, and $P$:
\begin{align}
r_1 &= \tfrac{1}{2}\sqrt{1 + \Sigma + T + P}, \hspace{-75pt}& r_2 &= \tfrac{1}{2}\sqrt{1 + \Sigma - T - P}\nonumber\\
r_3 &= \tfrac{1}{2}\sqrt{1 - \Sigma - T + P}, \hspace{-75pt}& r_4 &= \tfrac{1}{2}\sqrt{1 - \Sigma + T - P}.\label{eq:transversity_moduli}
\end{align}
This means that the four moduli of the normalized transversity amplitudes can be determined unambiguously from measurements of the three single asymmetries. Note, that only three of these moduli are independent as $r_1^2 + r_2^2 + r_3^2 + r_4^2 = 1$.

There are six possible combinations for the relative phases of the transversity amplitudes, namely $\alpha_{ij} = \alpha_i - \alpha_j$ $(i \neq j)$. However, only three of these are independent. By fixing a certain reference phase $\alpha_l$, the three independent phases are denoted by $\delta_i = \alpha_i - \alpha_l$ ($i \neq l)$ and the dependent phases by $\Delta_{ij} = \delta_i - \delta_j$ ($i \neq j$). A specific complete set, consisting of three single and four well-chosen double asymmetries, gives access to a specific set of two independent and two dependent phases: $\{\delta_i, \delta_j, \Delta_{ik}, \Delta_{jk}\}$ ($i \neq j \neq k$). Two kinds of complete sets can be distinguished: complete sets of the first kind which have four different solutions to the set $\{\delta_i, \delta_j, \Delta_{ik}, \Delta_{jk}\}$, and complete sets of the second kind which have eight different solutions. The actual solution satisfies the trivial relation $\delta_i - \Delta_{ik} - \delta_j + \Delta_{jk} = 0$.

\section{Results}
\label{sec:results}
\begin{figure}[t!]
\centering
\includegraphics[viewport = 55 745 574 433, scale = 0.75, clip = true]{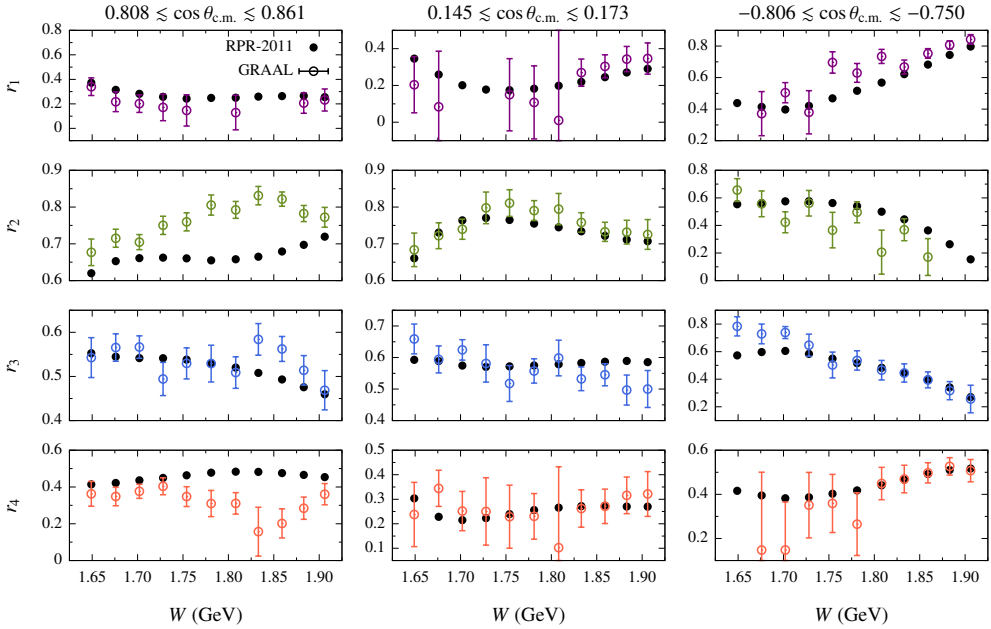}
\vspace{230pt}
\caption{(Color online) The moduli $r_{i}$ for the $\gamma p \to K^{+} \Lambda$ reaction as a function of the invariant mass $W$ for three $\cos\thcm$ ranges. The data are extracted from the GRAAL measurements of the single-polarization observables \cite{Lleres2007,Lleres2009}. The dots are the bin-center RPR-2011 predictions.}
\label{fig:moduli_GRAAL_RPR}
\end{figure}

Figure \ref{fig:moduli_GRAAL_RPR} shows the extracted transversity moduli $r_i$ from the GRAAL data for the $\{\Sigma, T, P\}$ of the $\gamma p \to K^{+} \Lambda$ reaction \cite{Lleres2007,Lleres2009} along with the corresponding predictions of the RPR-2011 model \cite{Corthals2006, DeCruz2012a, DeCruz2012b}. The results of Fig.~\ref{fig:moduli_GRAAL_RPR} indicate that with current experimental accuracies, knowledge about the three single asymmetries allows to determine the moduli $r_i$. It is seen that in a few kinematic situations one or more extracted moduli are missing. This is due to finite experimental resolution resulting a negative argument of one of the square roots in Eq.~\ref{eq:transversity_moduli}. To this day, there is not a single pseudoscalar-meson photoproduction reaction for which a complete data set has been published. The presented GRAAL data is the sole set of data which allows extracting the kinematic dependence of the moduli of the normalized transversity amplitudes.

In order to study the feasibility of extracting the relative phases from data, one has to resort to simulated of complete sets of observables. Here, the complete set $\{\Sigma,T,P;C_x,O_x,E,F\}$, which is of the first kind, will be considered for the $\gamma p \to K^{+} \Lambda$ reaction. From this set the phases $\{\delta_1, \delta_2, \Delta_{13}, \Delta_{23}\}$ (with $\alpha_4$ as the reference phase) can be obtained \cite{Vrancx:2013pza}. Measured asymmetries are simulated by generating events from a Gaussian distribution with the RPR-2011 prediction as the mean value and a standard deviation given by a certain experimental resolution $\sigexp$.

\begin{figure}[t!]
\centering
\includegraphics[viewport = 55 745 574 400, scale = 0.75, clip = true]{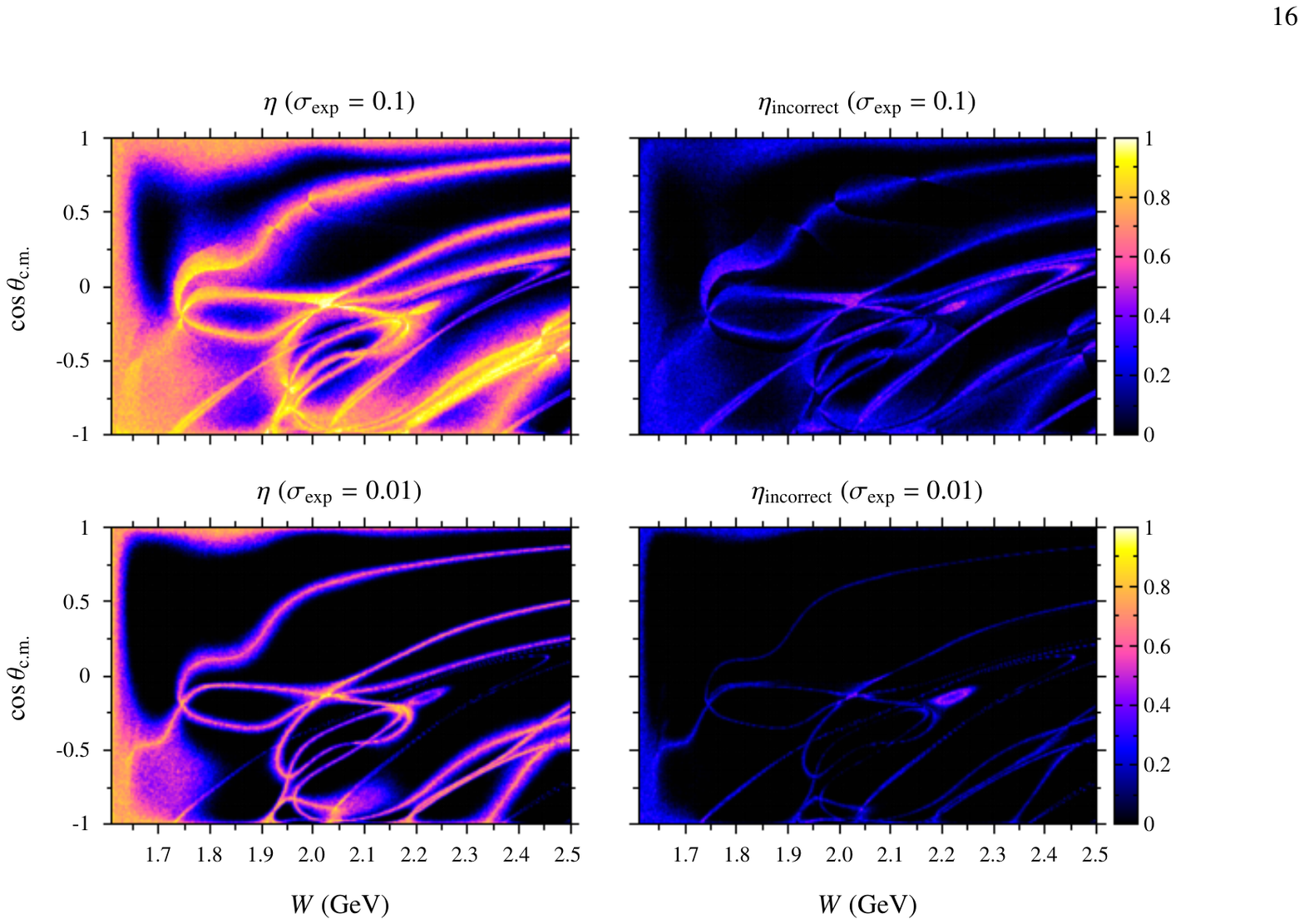}
\vspace{237pt}
\caption{(Color online) The $\{\Sigma, T, P; C_x,O_x,E,F\}$ insolvabilities $\eta = \eta_\text{imaginary} + \eta_\text{incorrect}$ and $\eta_\text{incorrect}$ as a function of $W$ and $\cos\thcm$ for two values of the input experimental resolution: $\sigexp = 0.1$ and $\sigexp = 0.01$. At each of the $90\,375$ kinematic points, $1000$ simulated data sets were generated with the RPR-2011 model, from which the average insolvabilities were calculated.}
\label{fig:insolvability_maps}
\end{figure}

When experimental uncertainty is involved, none of the four solutions to $\{\delta_1, \delta_2, \Delta_{13}, \Delta_{23}\}$ will satisfy the constraint $\delta_1 - \Delta_{13} - \delta_2 + \Delta_{23} = 0$ exactly in general. The most likely actual solution is the one for which the absolute value of the ratio of the evaluated constraint to its corresponding error is the smallest. However, it is possible that the most likely solution does not coincide with the actual solution. This can be readily verified by comparing the most likely solution with the RPR-2011 predictions that generated the simulated data. A solution that is identified as the most likely one, but is not the actual solution is dubbed an \emph{incorrect solution}. Another possibility, is the occurrence of imaginary solutions for one or more of the moduli and/or phases, as was apparent from Fig.~\ref{fig:moduli_GRAAL_RPR} for example. This type of solutions is referred to as \emph{imaginary solutions}.

The \emph{insolvability} $\eta(W, \cos\thcm)$ at a specific kinematic point is introduced as the fraction of simulated complete data sets that are solved incorrectly or have imaginary solutions: $\eta = \eta_{\textrm{incorrect}} + \eta_{\textrm{imaginary}}$. Figure \ref{fig:insolvability_maps} shows the $\{\Sigma, T, P; C_x,O_x,E,F\}$ insolvabilities $\eta$ and $\eta_{\textrm{incorrect}}$ for $\sigexp = 0.1$ and $\sigexp = 0.01$. It is seen that for $\sigexp = 0.1$ the insolvability can become quite substantial and that the largest contribution is attributed to imaginary solutions. Increasing the experimental resolution by a factor of ten clearly improves the overall solvability.

Although incorrect solutions make up the smaller contribution to $\eta$, they can never really be identified in an analysis of real data without invoking a model. Incorrect solutions originate from assigning the most likely solution as the actual solution, which is not a statistically sound procedure though. A more conservative approach would consist of imposing a tolerance level on the confidence interval of the most likely solution. Then, the most likely solution would only be `accepted' as the actual solution when it has a certain minimum statistical significance. As discussed at the end of Sec.~IV C 2 in Ref.~\cite{Vrancx:2013pza}, however, imposing a tolerance confidence level would not be effective as the entire elimination of the incorrect solutions would lead to a rejection of the lion's share of the fraction of correct solutions. This would result in an almost $100\%$ insolvability.

Summarizing, using real data it was illustrated that measurements of the single asymmetries allow to map the moduli of the normalized transversity amplitudes fairly well. Extracting the phases requires also double asymmetries. For infinite precision, four of those are required to form a complete set. For finite experimental resolution, it was shown that in a lot of situations the phases cannot be determined unambiguously. Therefore, theoretical completeness does not necessarily imply experimental completeness. It remains to be investigated whether ``overcomplete'' sets, containing additional double asymmetries, could help resolve the spurious phase ambiguities.

\begin{acknowledgement}
This work is supported by the Research Council of Ghent University and the Flemish Research Foundation (FWO Vlaanderen).
\end{acknowledgement}

\end{document}